\documentclass[oneside,final,11pt]{amsart}

\title[A Criticism of ``An experimental test of non-local realism'']%
{A Criticism of the article ``An experimental test of non-local realism''}

\author[D. V. Tausk]{Daniel V. Tausk}

\address{Departamento de Matem\'atica,\hfill\break\indent  Universidade de S\~ao Paulo, Brazil}
\email{tausk@ime.usp.br}
\urladdr{http://www.ime.usp.br/\~{}tausk}

\date{September 7th, 2008}

\newcommand{\dd}{\mathrm d}

\begin{document}


\begin{abstract}
I make some critical comments on the article \cite{Zei}. This article makes incorrect claims concerning
Bell's theorem \cite{Bell}. Moreover, I point to the fact that a hypothesis referred to as ``realism'' in \cite{Zei}
is not used in the deduction of Leggett's inequality.
\end{abstract}

\maketitle

\bigskip

\hbox to \hsize{\small\hfil\vbox{\hsize=220pt``{\em It is the requirement of locality, or more precisely that the result of a measurement on one system be unaffected
by operations on a distant system with which it has interacted in the past, that creates the essential difficulty.}''

\medskip

John Bell in \cite[Introduction]{Bell}.}}

\bigskip

There is hardly a result that is more widely misunderstood in the scientific community than Bell's theorem. In a nutshell, there is
a widespread belief that in his celebrated article \cite{Bell}, Bell has shown that the assumption of locality,
when taken together with an assumption of the existence of hidden variables, allows one to deduce a certain inequality
(now known as {\em Bell's inequality}) that contradicts quantum mechanical predictions. A crucial point that
is usually overlooked is the fact that the existence of the hidden variables used in the deduction of Bell's
inequality is {\em inferred from the assumption of locality\/} using the EPR argument;
it is not, as many physicists seem to think, an additional assumption that is necessary for proving the inequality.
Therefore, {\em violation of Bell's inequality implies that locality has to be abandoned}.

The misunderstanding regarding the meaning of Bell's theorem seems to be
widespread, but fortunately not universal. Indeed, attempts to clear up the confusion have been made
by many authors (see, for instance, \cite[pgs.\ 10---15]{Bricmont}, \cite{GoldBell}, \cite{Maudlin}, \cite{NorsenBell}),
most notably Bell himself. For instance, in footnote number 10 of \cite{BellSocks}, Bell wrote\footnote{%
Bell explains his violation of locality argument in several of his writings (see
\cite{BellSpeakable} for the collection of Bell's articles on the Foundations of Quantum Mechanics).
In fact, one version of the argument goes {\em directly\/} from locality to an inequality
that is violated by the quantum predictions (the Clauser--Holt--Shimony--Horne inequality) and it does not
even have to {\em mention\/} any sort of hidden variables (see, for instance, \cite{Bellcascade}
and \cite[Section 4]{BellSocks}). See also \cite{Bellcuisine} for the (in my opinion) best of Bell's articles on the subject.}:

\medskip

``{\em [\dots] My own first paper on this subject (\/{\em Physics \textbf 1, 195 (1965).}) starts with a
summary of the EPR argument \textbf{from locality to} deterministic hidden variables. But the commentators
have almost universally reported that it begins with deterministic hidden variables.}'' (emphasis in the original)

\medskip

The above noted misconception regarding the hypotheses necessary for the deduction of
Bell's inequality is not the only one; in addition, one encounters the now apparently popular habit of replacing
the expression ``hidden variables'' with the word ``realism''. This often leads
to a confusion between the notion of ``hidden variable'' (usually understood as any parameter that represents
more information about a given system than the wave function) with the several different possible meanings\footnote{%
See \cite{NorsenReal} for a nice discussion of some possible meanings of the word ``realism'' and for more detailed criticism
on the bad habit of saying that Bell has shown that we have to abandon ``local realism''. For a nice discussion about
the distinction between a {\em realist\/} and an {\em instrumentalist\/} interpretation of a given
physical theory, see \cite{SokBric}. There is indeed a relation between the notion of ``hidden variable'' and
some notions of ``realism'', but confusion often arises when one takes their meanings to be identical or when one simply does
not care to assign any precise meanings to those words at all.}
that can be given to the word ``realism'' in the Philosophy of Science (or Philosophy in general).
For instance, right at the beginning of paper \cite{Zei}, in the abstract, we find the following statement:

\bigskip

``{\em Most working scientists hold fast to the concept of `realism' --- a viewpoint according to which
an external reality exists independent of observation. But quantum physics has shattered some
of our cornerstone beliefs. According to Bell's theorem, any theory that is based on the joint
assumption of realism and locality (meaning that local events cannot be affected by actions in spacelike
separated regions) is at variance with certain quantum predictions.}''

\bigskip

So, the ``realism'' that appears in the abstract of \cite{Zei} seems to be some sort of philosophical notion of realism
(as we will see below, a different notion of ``realism'' is considered within the article). In fact, the citation
above looks like a suggestion that working scientists should take solipsism seriously, or at least that they should
consider the possibility that tables, mountains and planets do not exist at all in the absence of a conscious observer
of some sort. There is no rational justification for such a radically counter-intuitive statement.

A rather extreme type of misunderstanding concerning Bell's theorem is shown in the following passage from \cite[pg.\ 871]{Zei}:

\bigskip

``{\em The logical conclusion one can draw from the violation of local realism is that at least one of its assumptions
fails. Specifically, either locality or realism or both cannot provide a foundational basis for quantum theory. Each
of the resulting possible positions has strong supporters and opponents in the scientific community. However, Bell's
theorem is \textbf{unbiased} with respect to these views: on the basis of this theorem, one cannot, even in principle, favour
one over the other. It is therefore important to ask whether incompatibility theorems similar to Bell's can be found
in which at least one of these concepts is relaxed.}'' (emphasis added)

\bigskip

In this passage, the authors of \cite{Zei} give the impression that they think that the choice between
abandoning locality or abandoning realism (whatever that turns out to mean) is a matter of subjective personal taste: some
members of the scientific community prefer to abandon locality and others prefer to abandon realism. The statement
that Bell's theorem is ``unbiased'' reveals a great deal of misunderstanding regarding the hypotheses involved
in its proof.

Let us now turn to the analysis of the hypotheses that the authors of \cite{Zei} claim to be using in the proof
of Leggett's inequality (\cite[pg.\ 872]{Zei}):

\bigskip

``{\em [\dots] The theories are based on the following assumptions: (1) all measurement outcomes are determined
by pre-existing properties of particles independent of the measurement (realism); (2) physical states are statistical
mixtures of subensembles with definite polarization, where (3) polarization is defined such that expectation values
taken for each subensemble obey Malus' law (that is, the well-known cosine dependence of the intensity of a polarized
beam after an ideal polarizer).}''

\bigskip

The notion of ``realism'' expressed in hypothesis (1) is different from the one considered in the paper's abstract;
here ``realism'' doesn't seem to be taken as the opposite of solipsism or of some other radical form of philosophical idealism.
It is not completely clear, though, what hypothesis (1) means. Below I will present a brief analysis of some possible meanings
for hypothesis (1), but let me emphasize that my main point here is that {\em hypothesis (1) is simply not used in the deduction of Leggett's
inequality}. It is a bit odd, to say the least, that an article that claims to be doing ``an experimental test of non-local realism''
is apparently trying to accomplish its goal by verifying the violation of an inequality whose proof does not use
the very hypothesis that the authors call ``realism''! A more appropriate title would be ``An experimental test
of Leggett's subensembles hypothesis'', which is precisely what the article presents.

Let us discuss some possible meanings for hypothesis (1). What exactly is supposed to determine the outcome of
an experiment? One possible answer is: the outcome of an experiment is assumed to be completely determined
by a complete description of the system of particles being observed and by the self-adjoint operator that would normally
be used to compute the probability distribution on the set of possible outcomes for the experiment. I'll call such
a hypothesis {\em operator realism}. The problem with operator realism is that it is already well-known that it is
inconsistent with quantum mechanical predictions: all those results that are usually called ``no hidden variables theorems''
prove precisely that\footnote{%
See, for instance, \cite[pgs.\ 6---10]{Bricmont} and \cite{Mermin}. A notable exception is the ``no hidden variables theorem''
due to von Neumann, which does not even allow the rejection of operator realism, since (in addition to operator
realism) it assumes a hypothesis that is not justifiable solely on the basis of quantum mechanical predictions.
See \cite{Mermin} for details.}. Thus, it wouldn't make much sense to take hypothesis (1) as being operator realism.
An alternative meaning to hypothesis (1) is the following: the outcome of an experiment is assumed to be completely
determined by a complete description of the system of particles being observed and by a sufficiently complete
description of the experimental apparatus\footnote{%
In the case of experiments with entangled particles located far away from each other, the ``sufficiently
complete description of the experimental apparatus'' may include a description of parts of the experimental
apparatus that are located far away from each other.}. For the purposes of the present discussion, I'll call such a hypothesis
{\em determinism}. It is very important to understand that determinism is very different from operator realism. Indeed,
determinism {\em is\/} consistent with all quantum mechanical predictions: this is proven by the existence of
Bohmian Mechanics, which is a completely deterministic theory that agrees with all quantum mechanical predictions\footnote{%
See \cite{GoldBohm,Tumulka}. Bohmian Mechanics is a formalism that is obtained from the standard quantum formalism by
adding well-defined particle trajectories satisfying a simple first order ordinary differential equation involving
the wave function. The addition of particle trajectories turns quantum mechanics into a precisely formulated physical theory,
with a well-defined ontology. The privileged role played by ``measurement'' or ``observers'' in standard quantum
mechanics simply disappears; in particular, Bohmian Mechanics solves the so called ``measurement problem''.
Bohmian Mechanics is obviously incompatible with operator realism (see \cite{GoldNaive}).}.
Therefore, any attempt to prove incompatibility between determinism and quantum mechanical predictions is completely
pointless. It does make sense, however, to try to prove that determinism plus some additional hypotheses leads
to some inequality that contradicts quantum mechanical predictions.
Unfortunately, as it will be explained below, that is not what is done in \cite{Zei}:
hypothesis (1) (be it determinism or something else) is not used at all in the proof of Leggett's inequality.

I will now present a derivation of Leggett's inequality from hypotheses (2) and (3) of \cite{Zei}
alone. As usual, the experimenters will be called Alice and Bob. Let $A$ (resp., $B$) denote the result of
Alice's (resp., Bob's) experiment, which can be either $1$ or $-1$, and $\mathbf a$
(resp., $\mathbf b$) denote the setting of Alice's (resp., Bob's) experimental
apparatus, which is an element of the unit sphere $S^2$. Hypotheses (2) and (3) amount to a ``hidden variables''
assumption, namely, that vectors $\mathbf u,\mathbf v\in S^2$ are associated to each run of the experiment
(in addition to the usual quantum state vector). A precise mathematical statement of (2) and (3) can be formulated as
follows:

\medskip

\begin{itemize}
\item[($\bullet$)] for every $\mathbf a,\mathbf b\in S^2$, there exists a probability distribution $\mathbb P_{\mathbf{ab}}$ on
$\{-1,1\}\times\{-1,1\}\times S^2\times S^2$ representing the joint distribution of $(A,B,\mathbf u,\mathbf v)$ when
the experiment is done with settings $\mathbf a$ and $\mathbf b$. These distributions are
such that the (marginal) distribution $P_{\mathbf{uv}}$ of $(\mathbf u,\mathbf v)$ does not depend on $(\mathbf a,\mathbf b)$,
and the conditional expectations $E_{\mathbf{ab}}(A\mid\mathbf u,\mathbf v)$ and $E_{\mathbf{ab}}(B\mid\mathbf u,\mathbf v)$
are given by:
\[E_{\mathbf{ab}}(A\mid\mathbf u,\mathbf v)=\mathbf u\cdot\mathbf a,\quad
E_{\mathbf{ab}}(B\mid\mathbf u,\mathbf v)=\mathbf v\cdot\mathbf b.\leqno{(*)}\]
\end{itemize}

\medskip

Here is how ($\bullet$) leads to Leggett's inequality: equality (5) in \cite{Zei}, namely:
\[-1+\vert A+B\vert=AB=1-\vert A-B\vert,\]
is simply a consequence of the fact that $A$ and $B$ take values on $\{-1,1\}$. Taking expectations conditioned on
$(\mathbf u,\mathbf v)$ and using the fact that the modulus of the expectation is less than or equal to the
expectation of the modulus, we get:
\begin{multline*}
-1+\big\vert E_{\mathbf{ab}}(A\mid\mathbf u,\mathbf v)+E_{\mathbf{ab}}(B\mid\mathbf u,\mathbf v)\big\vert\le
E_{\mathbf{ab}}(AB\mid\mathbf u,\mathbf v)\\
\le1-\big\vert E_{\mathbf{ab}}(A\mid\mathbf u,\mathbf v)-E_{\mathbf{ab}}(B\mid\mathbf u,\mathbf v)\big\vert,
\end{multline*}
which is the same as inequality (8) in \cite{Zei}. We now use ($*$) and take expectations on both sides (with fixed
$(\mathbf a,\mathbf b)$), obtaining:
\begin{multline*}
-1+\int_{S^2\times S^2}\vert\mathbf u_0\cdot\mathbf a+\mathbf v_0\cdot\mathbf b\vert\;\dd P_{\mathbf{uv}}(\mathbf u_0,\mathbf v_0)
\le E_{\mathbf{ab}}(AB)\\
\le1-\int_{S^2\times S^2}\vert\mathbf u_0\cdot\mathbf a-\mathbf v_0\cdot\mathbf b\vert\;\dd P_{\mathbf{uv}}(\mathbf u_0,\mathbf v_0).
\end{multline*}
The rest of the proof of Leggett's inequality involves only mathematical
manipulations of the inequalities above and makes no further use of physical hypotheses of any kind.

\bigskip

To summarize, what can one conclude from the violation of Leggett's inequality? The logical conclusion is that
Leggett's subensembles hypothesis ($\bullet$) is false, i.e., that a theory that contains the hidden variables
$\mathbf u$ and $\mathbf v$ proposed by Leggett cannot be empirically viable. That doesn't tell us anything about
determinism or any type of philosophical realism. A title like ``An experimental test of non-local realism''
is severely misleading: it could, for instance, lead some readers into believing that the experiment reported by the article
makes a theory like Bohmian Mechanics more implausible\footnote{%
It should be observed, however, that the authors of \cite{Zei} are careful enough to point out in their article that Bohmian
Mechanics falls outside the class of theories that is falsified by their experiment. On the other hand, the authors
do not present any arguments against Bohmian Mechanics itself and I'm very puzzled about why the authors
make wild speculations about the violation of things like Aristotelian logic in the last paragraph of the article.},
while it is exactly the other way around: a prediction
of Bohmian Mechanics has been experimentally verified and a class of alternatives to it has been shown not to be
viable.

\end{document}